\documentclass[aps,showpacs,preprintnumbers,amsmath,amssymb,nofootinbib]{revtex4}
\usepackage{graphicx}

\begin{document}

\preprint{gr-qc/0401069}

\title{Anisotropy in Bianchi-type brane cosmologies}

\author{T. Harko}
\email{harko@hkucc.hku.hk} \affiliation{ Department of Physics,
The University of Hong Kong, Pokfulam Road, Hong Kong }

\author{M. K. Mak}
\email{mkmak@vtc.edu.hk} \affiliation{ Department of Physics, The
University of Hong Kong, Pokfulam Road, Hong Kong }

\date{January 16, 2004}


\begin{abstract}

The behavior near the initial state of the anisotropy parameter of
the arbitrary type, homogeneous and anisotropic Bianchi models is
considered in the framework of the brane world cosmological
models. The matter content on the brane is assumed to be an
isotropic perfect cosmological fluid obeying a barotropic equation
of state. To obtain the value of the anisotropy parameter at an
arbitrary moment an evolution equation is derived, describing the
dynamics of the anisotropy as a function of the volume scale
factor of the Universe. The general solution of this equation can
be obtained in an exact analytical form for the Bianchi I and V
types and in a closed form for all other homogeneous and
anisotropic geometries. The study of the values of the anisotropy
in the limit of small times shows that for all Bianchi type
space-times filled with a non-zero pressure cosmological fluid,
obeying a linear barotropic equation of state, the initial
singular state on the brane is isotropic. This result is obtained
by assuming that in the limit of small times the asymptotic
behavior of the scale factors is of Kasner-type. For brane worlds
filled with dust, the initial values of the anisotropy coincide in
both brane world and standard four-dimensional general
relativistic cosmologies.

\end{abstract}


\maketitle

\section{Introduction}

The idea \cite{RS99a} that our four-dimensional Universe might be a
three-brane, embedded in a higher dimensional space-time, has attracted much
attention. According to the brane-world scenario, the physical fields in our
four-dimensional space-time, which are assumed to arise as fluctuations of
branes in string theories, are confined to the three brane. Only gravity can
freely propagate in the bulk space-time, with the gravitational
self-couplings not significantly modified. This model originated from the
study of a single $3$-brane embedded in five dimensions, with the $5D$
metric given by $ds^{2}=e^{-f(y)}\eta _{\mu \nu }dx^{\mu }dx^{\nu }+dy^{2}$,
which, due to the appearance of the warp factor, could produce a large
hierarchy between the scale of particle physics and gravity. Even if the
fifth dimension is uncompactified, standard $4D$ gravity is reproduced on
the brane. Hence this model allows the presence of large, or even infinite
non-compact extra dimensions. Our brane is identified to a domain wall in a $%
5$-dimensional anti-de Sitter space-time.

The Randall-Sundrum model was inspired by superstring theory. The
ten-dimensional $E_{8}\times E_{8}$ heterotic string theory, which contains
the standard model of elementary particle, could be a promising candidate
for the description of the real Universe. This theory is connected with an
eleven-dimensional theory, M-theory, compactified on the orbifold $%
R^{10}\times S^{1}/Z_{2}$ \cite{HW96}. In this model we have two
separated ten-dimensional manifolds. For a review of dynamics and
geometry of brane Universes see \cite{Ma03}.

The static Randall-Sundrum solution has been extended to
time-dependent solutions and their cosmological properties have
been extensively studied \cite{KK00}-\cite{AgChLa03}. In one of
the first cosmological applications of this scenario, it was
pointed out that a model with a non-compact fifth dimension is
potentially viable, while the scenario which might solve the
hierarchy problem predicts a contracting Universe, leading to a
variety of cosmological problems \cite{CsGrKoTe99}. By adding
cosmological constants to the brane and bulk, the problem of the
correct behavior of the Hubble parameter on the brane has been
solved by Cline, Grojean and Servant \cite {ClGrSe99}. As a result
one also obtains normal expansion during nucleosynthesis, but
faster than normal expansion in the very early Universe. The
creation of a spherically symmetric brane-world in AdS bulk has
been considered, from a quantum cosmological point of view, with
the use of the Wheeler-de Witt equation, by Anchordoqui, Nunez and
Olsen \cite {AnNuOl00}.

The effective gravitational field equations on the brane world, in which all
the matter forces except gravity are confined on the $3$-brane in a $5$%
-dimensional space-time with $Z_{2}$-symmetry have been obtained, by using a
geometric approach, by Shiromizu, Maeda and Sasaki \cite{SMS00,SSM00}. The
correct signature for gravity is provided by the brane with positive
tension. If the bulk space-time is exactly anti-de Sitter, generically the
matter on the brane is required to be spatially homogeneous. The contraction
of the $5$-dimensional Weyl tensor with the normal to the brane $E_{IJ}$
gives the leading order corrections to the conventional Einstein equations
on the brane. The four-dimensional field equations for the induced metric
and scalar field on the world-volume of a $3$-brane in the five-dimensional
bulk, with Einstein gravity plus a self-interacting scalar field, have been
derived by Maeda and Wands \cite{MW00}.

Realistic brane-world cosmological models require the
consideration of more general matter sources to describe the
evolution and dynamics of the very early Universe. The influence
of the bulk viscosity of the matter on the brane has been
considered, for an isotropic flat Friedmann-Robertson-Walker (FRW)
geometry, in \cite{ChHaMa01b} and, for a Bianchi type I geometry,
in \cite{HaMa03}. The first order rotational perturbations of
isotropic FRW cosmological models have been studied in
\cite{ChHaKaMa02} and \cite{ChHaKaMa03}.

The general solution of the field equations for an anisotropic
brane with Bianchi type I and V geometry, with perfect fluid and
scalar fields as matter sources, has been obtained in
\cite{ChHaMa01a}. Expanding Bianchi type I and V brane-worlds
always isotropize. Anisotropic Bianchi type I brane-worlds with a
pure magnetic field and a perfect fluid have also been analyzed
\cite {BaHe01}. Limits on the initial anisotropy induced by the
5-dimensional Kaluza-Klein graviton stresses by using the CMB
anisotropies have been obtained by Barrow and Maartens
\cite{BaMa01}. The dynamics of a flat, isotropic brane Universe
with two-component matter source: a perfect fluid with a linear
barotropic equation of state and a scalar field with a power-law
potential has been investigated in \cite{SaTo03}. Solutions for
which the scalar field energy density scales as a power-law of the
scale factor (so called scaling solutions) have been obtained and
their stability analysis provided.

A family of Bianchi type braneworlds with anisotropy has been
constructed in \cite{Ca03}, by solving the five-dimensional field
equations in the bulk. The cosmological dynamics on the brane has
been analyzed by also including the Weyl term, and the relation
between the anisotropy on the brane and the Weyl curvature in the
bulk has been discussed. In these models, it is not possible to
achieve geometric anisotropy for a perfect fluid or scalar field
-- the junction conditions require anisotropic stresses on the
brane. But in an anti-de Sitter bulk, the solutions can isotropize
and approach a Friedmann type brane. Bianchi I type brane
cosmologies with scalar matter self-interacting through
combinations of exponential potentials have been studied in
\cite{AgLa03}. Such models correspond in some cases to
inflationary Universes. In the brane scenario, as happens in
standard four-dimensional general relativity, an increase in the
number of fields assists inflation. The asymptotic behavior of
Bianchi I brane worlds was considered in \cite{AgChLa03}. As a
consequence of the nonlocal anisotropic stresses induced by the
bulk, in the limit in which the mean radius goes to infinity the
brane does not isotropize and the nonlocal energy does not vanish.
The inflation due to the cosmological constant might be prevented
by the interaction with the bulk.

The study of anisotropic homogeneous brane world cosmological
models \cite {To01}-\cite{F01} has shown an important difference
between these models and standard four-dimensional general
relativity, namely, that brane Universes are born in an isotropic
state. For Bianchi type I and V geometries this type of behavior
has been found both by exactly solving the gravitational field
equations \cite{ChHaMa01a}, or from the qualitative analysis of
the model \cite{CS201}. A general analysis of the anisotropy in
spatially homogeneous brane world cosmological models has been
performed by Coley \cite{Co01a}, who has shown that the initial
singularity is isotropic, and hence the initial conditions problem
is naturally solved. Consequently, close to the initial
singularity, these models do not exhibit Mixmaster or chaotic-like
behavior \cite{Co01b}. Based on the results of the study of
homogeneous anisotropic cosmological models Coley \cite{Co01b}
conjectured that the isotropic singularity could be a general
feature of brane cosmologies. This conclusion has been contested
as a result of a perturbative analysis of the dimensionless shear
\cite{BrDu02}. The existence of a decaying mode in scalar
perturbations that grows unbounded in the past seems to suggest
that anisotropy also grows unbounded in the limit of small times
\cite{BrDu02}. This result is based on including, through
perturbations, generic inhomogeneities in the cosmological model
and these are responsible for the unbounded growth of the
anisotropy near the singularity. However, in a qualitative
numerical study of the asymptotic dynamical evolution of spatially
inhomogeneous brane-world cosmological models close to the initial
singularity, Coley, He and Lim \cite{CoHeLi03} have shown that
spatially inhomogeneous $G_{2}$ brane cosmological models with one
spatial degree of freedom always have an initial singularity,
which is characterized by the fact that spatial derivatives are
dynamically negligible. From the numerical analysis they have also
found that there is an initial isotropic singularity in all of
these spatially inhomogeneous brane cosmologies, including the
physically important cases of radiation and a scalar field source.
The numerical studies indicate that the singularity is isotropic
for all relevant initial conditions. A similar result has been
obtained by using the covariant and gauge-invariant approach for
the analysis of the linear perturbations of the isotropic model
${\cal F}_b$, which is a past attractor in the phase space of
homogeneous Bianchi models on the brane \cite{DuGoBrCo03}.
Therefore one can conclude that brane Universes are born with
naturally built-in isotropy, contrary to standard four-dimensional
general relativistic cosmology \cite{Co03}. The observed
large-scale homogeneity and isotropy of the Universe can therefore
be explained as a consequence of the initial conditions.

On the other hand most of the studies regarding the behavior of
the anisotropy in brane world cosmological models have been done
at a qualitative level, and have not provided explicit exact
representations for the anisotropy. Also many observationally
important questions, like, for example, the maximum value of the
anisotropy or the moment in the evolution of the Universe when
this maximum occurred have not been answered yet. It is also not
clear for what type of equation of state or range of parameters of
the cosmological fluid on the brane the initial singularity is
isotropic, or what is the effect of an inflationary phase on the
initial anisotropy.

It is the purpose of the present paper to consider the evolution
and dynamics of the anisotropy of homogeneous anisotropic (Bianchi
type) brane world cosmological models in a systematic way. As a
first step an evolution equation for the anisotropy parameter
(describing differences in the time expansion of the Universe
along the three principal axis) is derived. From mathematical
point of view it is a separable first order differential equation
for Bianchi types I and V, and a Bernoulli type equation for the
other Bianchi types. By integrating the evolution equation, one
can, generally, obtain the anisotropy parameter as a function of
the energy density and the volume scale factor of the Universe.
The study of the behavior of the anisotropy parameter near the
singular state shows that generally the initial value of this
parameter is dependent on the equation of state of the cosmic
matter. A high density cosmological fluid obeying a barotropic
equation of state starts its evolution on the brane from an
isotropic geometry, while the expansion of a pressureless dust, in
an anisotropic space-time, is similar to the standard
four-dimensional general relativistic one. But for inflationary
models the behavior of the anisotropy parameter is similar in both
brane world cosmological models and standard four-dimensional
general relativity.

The present paper is organized as follows. The basic equations of
the brane world cosmological models are presented in Section II.
The evolution of the anisotropy of Bianchi type I and V models is
considered in Section III. In Section IV the anisotropy of
arbitrary Bianchi type models is analyzed. Finally, in Section V,
we discuss and conclude our results.

\section{Gravitational field equations in the brane world model}

On the $5$-dimensional space-time (the bulk), with the negative vacuum
energy $\Lambda _{5}$ and brane energy-momentum as source of the
gravitational field, the Einstein field equations are given by
\begin{equation}
G_{IJ}=k_{5}^{2}T_{IJ},\qquad T_{IJ}=-\Lambda _{5}g_{IJ}+\delta (Y)\left[
-\lambda g_{IJ}+T_{IJ}^{\text{matter}}\right] ,
\end{equation}
In this space-time a brane is a fixed point of the $Z_{2}$ symmetry. In the
following capital Latin indices run in the range $0,...,4$, while Greek
indices take the values $0,...,3$.

Assuming a metric of the form $ds^{2}=(n_{I}n_{J}+g_{IJ})dx^{I}dx^{J}$, with
$n_{I}dx^{I}=d\chi $ the unit normal to the $\chi =\text{const.}$
hypersurfaces and $g_{IJ}$ the induced metric on $\chi =\text{const.}$
hypersurfaces, the effective four-dimensional gravitational equations on the
brane (the Gauss equation), take the form \cite{SMS00,SSM00}:
\begin{equation}
G_{\mu \nu }=-\Lambda g_{\mu \nu }+k_{4}^{2}T_{\mu \nu }+k_{5}^{4}S_{\mu \nu
}-E_{\mu \nu },  \label{Ein}
\end{equation}
where $S_{\mu \nu }$ is the local quadratic energy-momentum
correction,
\begin{equation}
S_{\mu \nu }=\frac{1}{12}TT_{\mu \nu }-\frac{1}{4}T_{\mu }{}^{\alpha }T_{\nu
\alpha }+\frac{1}{24}g_{\mu \nu }\left( 3T^{\alpha \beta }T_{\alpha \beta
}-T^{2}\right) ,
\end{equation}
and $E_{\mu \nu }$ is the non-local effect from the free bulk gravitational
field, the transmitted projection of the bulk Weyl tensor $C_{IAJB}$, $%
E_{IJ}=C_{IAJB}n^{A}n^{B}$, with the property $E_{IJ}\rightarrow E_{\mu \nu
}\delta _{I}^{\mu }\delta _{J}^{\nu }\quad $as$\quad \chi \rightarrow 0.$

The four-dimensional cosmological constant, $\Lambda $, and the coupling
constant, $k_{4}$, are given by $\Lambda =k_{5}^{2}\left( \Lambda
_{5}+k_{5}^{2}\lambda ^{2}/6\right) /2$ and $k_{4}^{2}=k_{5}^{4}\lambda /6$,
respectively, with $\lambda $ the vacuum energy on the brane.

The Einstein equation in the bulk and the Codazzi equation, also imply the
conservation of the energy momentum tensor of the matter on the brane, $%
D_{\nu }T_{\mu }{}^{\nu }=0$. Moreover, the contracted Bianchi identities on
the brane imply that the projected Weyl tensor should obey the constraint $%
D_{\nu }E_{\mu }{}^{\nu }=k_{5}^{4}D_{\nu }S_{\mu }{}^{\nu }$.

For any matter fields (scalar field, perfect or dissipative
fluids, kinetic gases etc.) the general form of the brane
energy-momentum tensor can be covariantly given as \cite{Ma96}
\begin{equation}
T_{\mu \nu }=\rho u_{\mu }u_{\nu }+ph_{\mu \nu }+\pi _{\mu \nu
}+2q_{(\mu }u_{\nu )}.  \label{EMT}
\end{equation}

The decomposition is irreducible for any chosen $4$-velocity $u^{\mu }$.
Here $\rho $ and $p$ are the energy density and isotropic pressure, and $%
h_{\mu \nu }=g_{\mu \nu }+u_{\mu }u_{\nu }$ projects orthogonal to $u^{\mu }$%
. The energy flux obeys $q_{\mu }=q_{<\mu >}$, and the anisotropic stress
obeys $\pi _{\mu \nu }=\pi _{<\mu \nu >}$, where angular brackets denote the
projected, symmetric and trace-free part:
\begin{equation}
V_{<\mu >}=h_{\mu }{}^{\nu }V_{\nu },\qquad W_{<\mu \nu >}=\left[ h_{(\mu
}{}^{\alpha }h_{\nu )}{}^{\beta }-\frac{1}{3}h^{\alpha \beta }h_{\mu \nu }%
\right] W_{\alpha \beta }.
\end{equation}

The symmetry properties of $E_{\mu \nu }$ imply that in general we can
decompose it irreducibly with respect to a chosen $4$-velocity field $u^{\mu
}$ as \cite{Ma00}
\begin{equation}
E_{\mu \nu }=-k^{4}\left[ {\cal U}\left( u_{\mu }u_{\nu }+\frac{1}{3}h_{\mu
\nu }\right) +{\cal P}_{\mu \nu }+2{\cal Q}_{(\mu }u_{\nu )}\right] ,
\label{WT}
\end{equation}
where $k=k_{5}/k_{4}$, ${\cal U}$ is a scalar, ${\cal Q}_{\mu }$ a
spatial vector and ${\cal P}_{\mu \nu }$ a spatial, symmetric and
trace-free tensor. For homogeneous models ${\cal Q}_{\mu }=0$ and
${\cal P}_{\mu \nu }=0$ \cite {CS01}. Hence the only non-zero
contribution from the 5-dimensional Weyl tensor from the bulk is
given by the scalar or ''dark radiation'' term ${\cal U} $.

In the following we shall consider only homogeneous and anisotropic brane
geometries. In particular, those which have space-like surfaces of
homogeneity (i.e. a $G_{3}$ acting simply transitively on a $V_{3}$). The
corresponding cosmological models fall into nine classes of equivalence, the
so-called Bianchi models. It is useful to classify the nine types into two
disjoint groups, depending on the different properties of the isometry
groups (the Lie groups), called class A and class B.

At each point of the space-time on the brane we take a local orthonormal
tetrad, where the metric can be written as
\begin{equation}
ds^{2}=-\left( \omega ^{0}\right) ^{2}+\sum_{i=1}^{3}\left( \omega
^{i}\right) ^{2},  \label{metric}
\end{equation}
where the differential one-forms $\omega ^{\mu }$ will be taken as $\omega
^{0}=dt$ and $\omega ^{i}=a_{i}(t)\Omega ^{i}$, where $\Omega ^{i}$ are
time-independent differential 1-forms and $a_{i}(t),i=1,2,3$ are the cosmic
scale factors. The time independent 1-forms obey the relations $d\Omega
^{i}=-\frac{1}{2}c_{kl}^{i}\Omega ^{k}\wedge \Omega ^{l}$, where the $%
c_{kl}^{i}$ are the canonical structure constants and $\wedge $
denotes the exterior product \cite{CaFa79}. Although the Ansatz
for the metric given by Eq. (\ref{metric}) is not the most general
one that one can chose, it is sufficient to display all the main
features of the behavior of Bianchi geometries.

With the help of the scale factors one can define the following variables: $%
V=\prod_{i=1}^{3}a_{i}$ \ (volume scale factor), $H_{i}=\dot{a}%
_{i}/a_{i},i=1,2,3$ (directional Hubble parameters), $H=\left( 1/3\right)
\sum_{i=1}^{3}H_{i}$ (mean Hubble parameter) and $\Delta H_{i}=H_{i}-H,\quad
i=1,2,3$. By using the definitions of $H$ and $V$ we immediately obtain $H=%
\dot{V}/3V$, where a dot denotes the derivative with respect to the
cosmological time $t$.

According to the definition of the energy-momentum tensor on the brane, Eq. (%
\ref{EMT}), in the general case of Bianchi type geometries, the
symmetry of the space-time allows different spatial components of
$T_{\mu \nu }$. There are several physical processes that could
generate an anisotropic energy momentum tensor, with
$T_{1}^{1}\neq T_{2}^{2}\neq T_{3}^{3}$, like magnetic fields,
heat transfer and/or viscous dissipative processes in the
cosmological fluid on the brane. The most important of these
processes are the bulk viscous type dissipative processes, which
are the main sources of entropy generation in the early Universe.
However,  the effect of the bulk viscosity of the cosmological
fluid on the brane can be
considered by adding to the usual thermodynamic pressure $p$ the bulk viscous pressure $%
\Pi $ and formally substituting the pressure terms in the
energy-momentum tensor by $p_{eff}=p+\Pi $ \cite{HaMa03}.
Therefore the consideration of the bulk viscosity of the
cosmological fluid does not lead to an anisotropic pressure
distribution.  The viscous dissipative anisotropic stress $\pi
_{\mu \nu }$ of the matter on the brane satisfies the evolution
equation $\tau _{2}h_{\alpha }^{\beta }h_{\beta }^{\nu
}\dot{\pi}_{\mu \nu }+\pi _{\alpha \beta }=-2\eta \sigma _{\alpha
\beta }-\left[ \eta T\left( \tau _{2}u^{\nu }/2\eta T\right)
_{;\nu }\pi _{\alpha \beta }\right] $ \cite{Ma96}, where $\eta $
is the shear viscosity coefficient, $\tau _{2}=2\eta \beta _{2}$,
with $\beta _{2}$ the thermodynamic coefficient for the tensor
dissipative contribution to the entropy density and $\sigma
_{\alpha \beta }$ is the shear tensor. Generally, it is assumed
that the dissipative contribution from the shear viscosity in the
early Universe can be neglected, $\eta \approx 0$ \cite{Ma96}.
Consequently, in the followings we consider that the anisotropic
stresses of the matter on the brane also vanish, $\pi _{\mu \nu
}\approx 0$. We suppose that in Eq. (\ref{EMT}) the heat transfer
is zero, that is, we take $q_{\mu }=0$. All these approximations
are standard in the analysis of the physics of the very early
Universe. Therefore in the following we assume that the pressure
distribution of the  cosmological fluid on the brane is isotropic
and the fluid pressure satisfies a barotropic equation of state of
the form $p=p\left( \rho \right) $.

For any homogeneous model the conservation equations of the energy
density of the matter $\rho $ on the brane and of the dark
radiation ${\cal U}$ can be written as
\begin{equation}
\dot{\rho}+3\left( \rho +p\right) H=0,  \label{cons}
\end{equation}
\begin{equation}
\dot{\mathcal{U}}+4H\mathcal{U}=0,
\end{equation}
leading to a general dependence of $\rho $ and $\mathcal{U}$ of
$V$ of the form
\begin{equation}
V=\frac{C_{0}}{w},\mathcal{U}=\frac{\mathcal{U}_{0}}{V^{4/3}},
\end{equation}
where
\begin{equation}
w=\exp \left[ \int \frac{d\rho }{\rho +p\left( \rho \right)
}\right]
\end{equation}
and $C_{0}\geq 0$ and $\mathcal{U}_{0}\geq 0$ are constants of
integration.

The modified Einstein gravitational field equations on the brane
can be written in the form of the standard Einstein
four-dimensional field equations,
\begin{equation}
G_{\mu \nu }=k_{4}^{2}T_{\mu \nu }^{(eff)},
\end{equation}
where $T_{\mu \nu }^{(eff)}=-\Lambda g_{\mu \nu }/k_4^2+T_{\mu \nu
}+k_{5}^{4}S_{\mu \nu }-\left( 1/k_{4}^{2}E_{\mu \nu }\right)
$\cite{Ma00}. Then the effective total energy density, pressure,
anisotropic stress and energy flux for a perfect fluid are $\rho
^{(eff)}=\Lambda /k_4^2 +\rho \left( 1+\rho /2\lambda \right)
+\left( 6/k_{4}^{4}\lambda \right) \mathcal{U}$,
$p^{(eff)}=p-\Lambda /k_4^2+\left( \rho /2\lambda \right) \left(
\rho +2p\right) +\left( 2/k_{4}^{4}\lambda \right) \mathcal{U}
$, $\pi _{\mu \nu }^{(eff)}=\left( 6/k_{4}^{4}\lambda \right) \mathcal{P}%
_{\mu \nu }$ and $q_{\mu }^{(eff)}=\left( 6k_{4}^{4}\lambda \right) \mathcal{%
Q}_{\mu }$. Since for homogeneous cosmological models $\mathcal{Q}_{\mu }=%
\mathcal{P}_{\mu \nu }=0$, it follows that in the case of a
perfect cosmological fluid there is a close analogy between the
gravitational field equations on the brane and standard
four-dimensional general relativity, with the role of the standard
energy density and pressure played by $\rho ^{(eff)}$ and
$p^{(eff)}$, respectively. The formal analogy between standard
four-dimensional general relativity and brane world cosmology
allows the immediate extension of the Collins-Hawking definition
of isotropization of a cosmological model \cite{CoHa73} to the
case of brane Universes.

Hence, we say that a brane world cosmological model approaches
isotropy if the following four conditions hold as $t\rightarrow
\infty $: i) the Universe is
expanding indefinitely and $H>0$ ii) $T^{(eff)00}>0$ and $T^{(eff)0i}/T^{(eff)00}%
\rightarrow 0$, $i=1,2,3$. $T^{(eff)0i}/T^{(eff)00}$ represents an
average velocity of the matter on the brane relative to the
surfaces of homogeneity. If this does not tend to zero, the
Universe would not appear homogeneous or isotropic iii) the
anisotropy in the locally measured Hubble constant $\sigma /H$
tends to zero, $\sigma /H\rightarrow 0$ and iv) the distortion
part of the metric tends to a constant. In condition iii) $\sigma
^2=(1/2)\sigma _{\mu \nu }\sigma ^{\mu \nu }$ represents the shear
of the normals $n_{\mu }$. For a metric of
the form $ds^{2}=dt^{2}-\exp \left( 2\alpha \right) \left[ \exp \left( 2\beta \right) %
\right] _{\mu \nu }\Omega ^{\mu }\Omega ^{\nu }$, where $\Omega
^{\mu }$ are one-forms that are not exact in general, $\alpha $ is
a time dependent function and $\beta $ is a symmetric traceless
matrix, the shear tensor is defined as $\sigma _{\mu \nu }=\left[
\exp \left( \beta \right) \right]
_{\lambda \mu }^{.}\left[ \exp \left( -\beta \right) \right] _{\lambda \nu }+%
\left[ \exp \left( \beta \right) \right] _{\lambda \nu }^{.}\left[
\exp \left( -\beta \right) \right] _{\lambda \mu }$ \cite{CoHa73}.
For Bianchi class A and B models the shear is given by $\sigma
^{2}=\sigma _{\mu \nu}\sigma ^{\mu \nu}/2=(1/2)\left(
\sum_{i=1}^{3}H_{i}^{2}-3H^{2}\right)$.

Spatially homogeneous models can be divided in three classes:
those which have less than the escape velocity (i.e., those whose
rate of expansion is insufficient to prevent them from
recollapsing), those which have just the escape velocity and those
which have more than the escape velocity \cite{CoHa73}. Models of
the third class do not tend, generally, to isotropy. In fact the
only types which can tend toward isotropy at arbitrarily large
times are types $I$, $V$, $VII_{0}$ and $VII_{h}$. For type
$VII_{h}$ there is no nonzero measure set of these models which
tends to isotropy \cite{CoHa73}. The Bianchi types that drive flat
and open
Universes away from isotropy in the Collins-Hawking sense are those of type $%
VII$.

As an indicator of the degree of anisotropy of a cosmological model one can
take the mean anisotropy parameter, defined according to \cite{CaFa79}
\begin{equation}
A=\frac{1}{3}\sum_{i=1}^{3}\left( \frac{\Delta H_{i}}{H}\right) ^{2}.
\label{an}
\end{equation}

For an isotropic cosmological model $H_{1}=H_{2}=H_{3}=H$ and
$A\equiv 0$. The anisotropy parameter is an important indicator of
the behavior of anisotropic cosmological models, since in standard
four-dimensional general relativity it is finite even for singular
states (for example, $A=2$ for Kasner-type geometries
\cite{ChHaMa01b}). The time evolution of $A$ is a good indicator
of the dynamics of the anisotropy.

For a homogeneous brane world model filled with a perfect fluid
satisfying a barotropic equation of state the conditions i), ii)
and iv) of the Collins-Hawking definition of isotropization are
naturally satisfied. Hence, if we consider only expanding brane
worlds, $H>0$ and condition i) holds. From the choice of the
matter content, and due to the symmetries of the energy-momentum
tensor, we have $T^{(eff)00}\equiv \rho^{(eff)}>0$ and
$T^{(eff)0i}\equiv 0$. The choice of the geometry implies that
condition iv) is also satisfied. Therefore, the value of the
parameter $\sigma /H$ is the main indicator of the
isotropic/anisotropic behavior of a brane world cosmological
model. As one can see from Eq. (\ref{an}) the quantity $\sigma /H$
is proportional to the square root of the anisotropy parameter,
$\sigma /H\sim \sqrt{A}$ and so, according to the Collins-Hawking
definition, if $A\rightarrow 0$, a brane world cosmological model
will isotropize in the large time limit $t\rightarrow \infty $.

The formal mathematical similarity between standard
four-dimensional general relativity and brane world theory can be
also used to extend the Hawking-Penrose singularity theorem
\cite{ElHa73} to brane cosmologies. From the definition of the
effective energy density $\rho ^{(eff)}$ and of the effective
pressure $p^{(eff)}$ it follows that for a linear barotropic fluid
with $p=(\gamma -1)\rho $, $1\leq \gamma \leq 2$, the effective
energy-momentum tensor on the brane satisfies both the strong and
weak energy conditions, which can be expressed as $\left(
T^{(eff)\mu \nu }-\left( 1/2\right) g^{\mu \nu }T^{(eff)}\right)
u_{\mu }u_{\nu }\geq 0$ and $T^{(eff)\mu \nu }u_{\mu }u_{\nu }\geq
0$, respectively, where $u_{\mu }$ is an arbitrary timelike
four-vector. The first of these conditions implies that the sum of
the local energy density and pressure is non-negative, $\rho
^{(eff)}+p^{(eff)}\geq 0$ and $\rho ^{(eff)}+3p^{(eff)}\geq 0$.
The second condition requires that the local energy density be
non-negative in every observer's rest frame, $\rho ^{(eff)}\geq 0$
and $\rho ^{(eff)}+p^{(eff)}\geq 0$. Therefore the Bianchi-type
brane spacetimes filled with a perfect linear barotropic fluid are
singular, since they satisfy the following conditions: a) $R_{\mu
\nu }u^{\mu }u^{\nu }\geq 0$ for all timelike vectors $u^{\mu }$
b) $u_{[\mu }R_{\nu ]\lambda \varepsilon \lbrack \sigma }u_{\rho
]}u^{\lambda }u^{\varepsilon }\neq 0$ for a vector $u^{\mu }$
tangent to some geodesic c) there are no closed time-like curves
and d) either i) there is a closed trapped surface or ii) there is
a point $p$ for which $u_{;\mu }^{\mu }<0$ for all of the vectors
$u^{\mu }$ tangent to the past light cone of $p$ \cite{ElHa73}.

\section{Evolution of the anisotropy in Bianchi type I and V models}

For a better understanding of the dynamics of the anisotropy in the brane
world we consider first in detail the evolution of the mean anisotropy
parameter $A$ in Bianchi type I and V geometries. From a formal point of
view these two geometries are described by the line element
\begin{equation}
ds^{2}=-dt^{2}+a_{1}^{2}(t)dx^{2}+a_{2}^{2}(t)e^{-2\alpha
x}dy^{2}+a_{3}^{2}(t)e^{-2\alpha x}dz^{2}.
\end{equation}

The metric for the Bianchi type I geometry formally corresponds to the case $%
\alpha =0$, while for the Bianchi type V case we have $\alpha =1$.

To study the time behavior of the anisotropy parameter $A$ near
the initial singular point on the brane we need
only the Einstein field equations involving the time derivatives of $%
H_{i},i=1,2,3$, and which are given by
\begin{equation}  \label{dVHi}
\frac{1}{V}\frac{d}{dt}(VH_{i})=\Lambda +\frac{2\alpha ^{2}}{V^{2/3}}+\,%
\frac{k_{4}^{2}}{2}\left( \rho -p\right) -\,\frac{k_{5}^{4}}{12}\rho p+\frac{%
1}{3}k^{4}\frac{\mathcal{U}_{0}}{V^{4/3}},\quad i=1,2,3.
\end{equation}

For $\alpha =0$ we obtain the $(ii),i\neq 0$ field equations for Bianchi
type I geometry, while $\alpha =1$ gives the Bianchi type V equations on the
brane world.

By summing Eqs. (\ref{dVHi}) we find
\begin{equation}
\frac{1}{V}\frac{d}{dt}(VH)=\dot{H}+3H^{2}=\Lambda +\frac{2\alpha ^{2}}{V^{2/3}}+\,%
\frac{k_{4}^{2}}{2}\left( \rho -p\right) -\frac{k_{5}^{4}}{12}\rho p+\frac{%
1}{3}k^{4}\frac{\mathcal{U}_{0}}{V^{4/3}}.  \label{dVH}
\end{equation}

Subtraction of Eq. (\ref{dVH}) from Eqs. (\ref{dVHi}) gives
\begin{equation}
\Delta H_{i}=H_{i}-H=\frac{K_{i}}{V},\qquad i=1,2,3,  \label{HiH}
\end{equation}
with $K_{i},\,i=1,2,3$ constants of integration satisfying the consistency
condition
\begin{equation}
\sum_{i=1}^{3}K_{i}=0.
\end{equation}

With the use of Eq. (\ref{HiH}) the anisotropy parameter defined
in Eq. (\ref {an}) becomes
\begin{equation}
A=\frac{K^{2}}{V^{2}H^{2}}=\frac{9K^{2}}{\dot{V}^{2}},
\label{an1}
\end{equation}
where $K^{2}=(1/3)\sum_{i=1}^{3}K_{i}^{2}$. Taking the time
derivative of Eq. (\ref{an1}) we obtain the following evolution
equation for the anisotropy:
\begin{equation}
\frac{dA}{dt}=-2\frac{\dot{H}+3H^{2}}{H}A,
\end{equation}
or, equivalently,
\begin{equation}
\frac{1}{A^{2}}\frac{dA}{dV}=-\frac{2}{3K^{2}}V\left(
\dot{H}+3H^{2}\right)
=-\frac{2}{3K^{2}}V\left[ \Lambda +\frac{2\alpha ^{2}}{V^{2/3}}+\frac{%
k_{4}^{2}}{2}\left( \rho -p\right) -\frac{k_{5}^{4}}{12}\rho p+\frac{1}{3}%
k^{4}\frac{\mathcal{U}_{0}}{V^{4/3}}\right] .  \label{an2}
\end{equation}

Taking into account that $d\rho /dV=-\left( \rho +p\right) /V$,
Eq. (\ref {an2}) can be written as
\begin{equation}
\frac{1}{A^{2}}\frac{dA}{d\rho
}=\frac{2C_{0}^{2}}{3K^{2}}\frac{1}{\left(
\rho +p\right) w^{2}}\left[ \Lambda +2\alpha ^{2}C_{0}^{-2/3}w^{2/3}+\,\frac{%
k_{4}^{2}}{2}\left( \rho -p\right) -\frac{k_{5}^{4}}{12}\rho p+\frac{1}{3}%
k^{4}\mathcal{U}_{0}C_{0}^{-4/3}w^{4/3}\right] ,
\end{equation}
with the general solution given by
\begin{equation}\label{an3a}
A\left( \rho \right) =-\frac{3K^{2}}{2C_{0}^{2} \int \left[
\Lambda +2\alpha
^{2}C_{0}^{-2/3}w^{2/3}+\,\frac{k_{4}^{2}}{2}\left( \rho -p\right)
-\,\frac{k_{5}^{4}}{12}\rho p+\frac{1}{3}k^{4}\mathcal{U}%
_{0}C_{0}^{-4/3}w^{4/3}\right] \left[ \left( \rho +p\right)
w^{2}\right] ^{-1}d\rho  -C},
\end{equation}
where $C$ is an arbitrary integration constant.

Eq. (\ref{an3a}) gives the general representation of the
anisotropy parameter as a function of the energy density of the
cosmological fluid for the Bianchi type I and V space-times in the
brane world scenario. For a cosmological fluid for which the
thermodynamic pressure $p$ obeys a linear
barotropic equation of state of the form  $p=(\gamma -1)\rho $, $\gamma =%
\text{const.}$, $1\leq \gamma \leq 2$, we have $\rho =\rho
_0/V^{\gamma }$, with $\rho _0 \geq 0$ a constant of integration.
$\rho _0$ can be expressed in terms of $C_0$ as $\rho
_0=C_0^{\gamma }$. Hence the anisotropy equation Eq. (\ref {an3a})
can be immediately integrated to give the general exact dependence
of the anisotropy parameter on the volume scale factor for Bianchi
type I and V geometries:
\begin{equation}
A\left( V\right) =\frac{3K^{2}}{\Lambda V^{2}+3\alpha
^{2}V^{4/3}+k_{4}^{2}\rho _{0}V^{2-\gamma
}+\frac{k_{5}^{4}}{12}\rho _{0}^{2}V^{2-2\gamma
}+k^{4}\mathcal{U}_{0}V^{2/3}+C},  \label{an3}
\end{equation}
where the arbitrary integration constant $C\neq 0$ is related, via
the field equations, to the constant $K^{2}$ by the relation
$C=2K^{2}/3$ \cite {ChHaMa01a}. The singular state at $t=0$ is
characterized by the condition $V(0)=0$. The value of the
anisotropy parameter for $t=0$ depends on the equation of state of
the cosmological fluid. Hence for $1<\gamma \leq 2$, from Eq.
(\ref{an3}) it follows
\begin{equation}
\lim_{V\rightarrow 0}A\left( V\right) =0,1<\gamma \leq 2.
\end{equation}

Therefore the singular state of the high density Bianchi type I and V brane
cosmological models is isotropic, with $A(0)=0$. For the case of the
pressureless dust filled anisotropic brane Universes, $p=0$ and $\gamma =1$.
In this case
\begin{equation}
\lim_{V\rightarrow 0}A\left( V\right)
=\frac{36K^{2}}{k_{5}^{4}\rho _{0}^{2}+12C},\gamma =1.
\end{equation}

The singular state of the dust filled brane Universe is anisotropic, with $%
A(0)\neq 0$. In the case of the standard four-dimensional general
relativity (SGR), the behavior of the anisotropy parameter is
different from the case of brane
cosmological models. SGR is recovered if the limits $k_{5}\rightarrow 0$, $%
k\rightarrow 0$ and $\Lambda _5\rightarrow -\infty$ are taken
simultaneously. Therefore the anisotropy parameter in standard
four-dimensional general relativity $A^{SGR}$ is given by
\begin{equation}
A^{SGR}\left( V\right) =\frac{3K^{2}}{\left( \Lambda V^{2}+3\alpha
^{2}V^{4/3}+k_{4}^{2}\rho _{0}V^{2-\gamma }+C\right) }.
\end{equation}

Near the singular state,
\begin{equation}
\lim_{V\rightarrow 0}A^{SGR}\left( V\right) =\frac{3K^{2}}{C}>0,\forall
\gamma \in \left[ 1,2\right] .
\end{equation}

In fact, one can show that for barotropic matter filled standard
four-dimensional general relativistic Bianchi type I and V models
$A^{SGR}\left( 0\right) \leq 2$ \cite{ChHaMa01a}.

Therefore in the case $\gamma =1$ we obtain the following relation between
the initial values of the anisotropy parameters $A$ and $A^{SGR}$ of the
brane world models and of the SGR, respectively:
\begin{equation}
A(0)=\frac{A^{SGR}(0)}{1+\frac{k_{5}^{4}\rho _{0}^{2}}{12C}}.
\end{equation}

In Bianchi type I and V geometries there is also a simple
proportionality relation between the shear scalar $\sigma ^{2}$
and the anisotropy parameter:
\begin{equation}
\sigma ^{2}=\sigma _{ik}\sigma ^{ik}/2=\frac{1}{2}\left(
\sum_{i=1}^{3}H_{i}^{2}-3H^{2}\right) =\frac{3}{2}AH^{2}.
\end{equation}

\section{Behavior of anisotropy in arbitrary type Bianchi models}

For arbitrary type Bianchi geometries the $\left( ii\right) ,i=1,2,3$
components of the gravitational field equations on the brane can be
represented in the following general form:
\begin{equation}\label{bian}
\frac{1}{V}\frac{d}{dt}(VH_{i})=F_{i}\left(
a_{1},a_{2},a_{3}\right)+ \Lambda +
\frac{k_{4}^{2}}{2}\left( \rho -p\right) -\frac{k_{5}^{4}}{12}\rho p+\frac{%
1}{3}k^{4}\frac{\mathcal{U}_{0}}{V^{4/3}},\quad i=1,2,3,
\end{equation}
where $F_{i}\left( a_{1},a_{2},a_{3}\right) ,i=1,2,3$ are functions which
depend on the Bianchi type. For the class A models \cite{JeSt86}
\begin{equation}
F_{1}\left( a_{1},a_{2},a_{3}\right) =\frac{\left(
c_{2}a_{2}^{2}-c_{3}a_{3}^{2}\right) ^{2}-\left( c_{1}a_{1}^{2}\right) ^{2}}{%
2V^{2}},
\end{equation}
where the constants $c_{i},i=1,2,3$ define the Bianchi type, and
are given in the table.

\begin{table}[h]
\begin{center}
\begin{tabular}{|c|c|c|c|}\hline
Bianchi type & $c_{1}$ & $c_{2}$ & $c_{3}$ \\\hline I & 0 & 0 &
0\\\hline II & 1 & 0 & 0 \\\hline VI$_{0}$ & 1 & -1 & 0 \\\hline
VII$_{0}$ & 1 & 1 & 0 \\ \hline VIII & 1 & 1 & -1 \\ \hline IX & 1
& 1 & 1 \\ \hline
\end{tabular}
\end{center}
\caption{Values of the constants $c_{i},i=1,2,3$, for the Bianchi
type A models \cite{CoHa73}, \cite{JeSt86}.}
\end{table}

$F_{2}\left( a_{1},a_{2},a_{3}\right) $ and $F_{3}\left(
a_{1},a_{2},a_{3}\right) $ can be obtained by a cyclic permutation of the
elements in the numerator of $F_{1}$. For Bianchi types V and VI$_{h}$
\begin{equation}
F_{1}\left( a_{1},a_{2},a_{3}\right) =-2\frac{a_{0}^{2}+q_{0}^{2}}{a_{1}^{2}}%
+2\frac{b^{2}}{a_{2}^{4}a_{3}^{2}},
\end{equation}
\begin{equation}
F_{2}\left( a_{1},a_{2},a_{3}\right) =-2\frac{a_{0}^{2}+a_{0}q_{0}}{a_{1}^{2}%
}-2\frac{b^{2}}{a_{2}^{4}a_{3}^{2}},
\end{equation}
\begin{equation}
F_{3}\left( a_{1},a_{2},a_{3}\right) =-2\frac{a_{0}^{2}-a_{0}q_{0}}{a_{1}^{2}%
},
\end{equation}
with $a_{0},q_{0}$ and $b$ constants.

If we take $q_{0}=b=0$ we obtain Bianchi type V, for $q_{0},b\neq 0$ we
obtain Bianchi type \ VI$_{h}$ ($h\neq 0$), while for $q_{0}=-1$ we have
Bianchi type III \cite{JeSt86}.

For Bianchi types IV and VII$_{h}$ ($h\neq 0$) the functions $F_{i}$ are
slightly more complicated. Thus for Bianchi type IV
\begin{equation}
F_{1}\left( a_{1},a_{2},a_{3}\right) =\frac{2}{a_{1}^{2}}+\frac{a_{3}^{2}}{%
2a_{1}^{2}a_{2}^{2}},
\end{equation}
\begin{equation}
F_{2}\left( a_{1},a_{2},a_{3}\right) =\frac{2}{a_{1}^{2}}+\frac{a_{3}^{2}}{%
2a_{1}^{2}a_{2}^{2}}+\frac{a_{3}^{2}}{2a_{2}^{2}}\dot{f}^{2},
\end{equation}
\begin{equation}
F_{3}\left( a_{1},a_{2},a_{3}\right) =\frac{2}{a_{1}^{2}}-\frac{a_{3}^{2}}{%
2a_{1}^{2}a_{2}^{2}}+\frac{a_{3}^{2}}{2a_{2}^{2}}\dot{f}^{2},
\end{equation}
with $f$ corresponding to the off-diagonal term \cite{HaTs77}. For Bianchi
type VII$_{h}$ ($h\neq 0$)
\begin{equation}
F_{1}\left( a_{1},a_{2},a_{3}\right) =2\Delta ^{2}a_{1}^{2}a_{2}^{2}\frac{%
a_{1}^{2}+a_{2}^{2}}{a_{1}^{2}-a_{2}^{2}}\left( 2\frac{\dot{a}_{3}}{a_{3}}-%
\frac{\dot{a}_{1}}{a_{1}}-\frac{\dot{a}_{2}}{a_{2}}\right) ,
\end{equation}
\begin{equation}
F_{2}\left( a_{1},a_{2},a_{3}\right) =-F_{1}\left( a_{1},a_{2},a_{3}\right) ,
\end{equation}
\begin{equation}
F_{3}\left( a_{1},a_{2},a_{3}\right) \equiv 0,
\end{equation}
where $\Delta =$constant$\neq 0$ \cite{BoMu78}.

Generally there is one more field equation, the $(00)$ equation, but it will
not be used in the proof of the results.

By adding Eqs. (\ref{bian}) on obtains
\begin{equation}
\frac{1}{V}\frac{d}{dt}(VH)=\dot{H}+3H^{2}=F\left(
a_{1},a_{2},a_{3}\right) +\Lambda +%
\frac{k_{4}^{2}}{2}\left( \rho -p\right) -\frac{k_{5}^{4}}{12}\rho p+\frac{%
1}{3}k^{4}\frac{\mathcal{U}_{0}}{V^{4/3}},  \label{bian1}
\end{equation}
where
\begin{equation}
F\left( a_{1},a_{2},a_{3}\right) =\frac{1}{3}\sum_{i=1}^{3}F_{i}\left(
a_{1},a_{2},a_{3}\right) .
\end{equation}

Substraction of Eq. (\ref{bian1}) from Eq. (\ref{bian}) and integration of
the resulting equation gives
\begin{equation}
\Delta H_{i}=H_{i}-H=\frac{K_{i}}{V}+\frac{1}{3V}\int \frac{\Delta F_{i}%
\left[ a_{1}\left( V\right) ,a_{2}\left( V\right) ,a_{3}\left( V\right) %
\right] }{H}dV,i=1,2,3,
\end{equation}
where
\begin{equation}
\Delta F_{i}\left( a_{1},a_{2},a_{3}\right) =F_{i}\left(
a_{1},a_{2},a_{3}\right) -F\left( a_{1},a_{2},a_{3}\right) ,i=1,2,3,
\end{equation}
with the property
\begin{equation}
\sum_{i=1}^{3}\Delta F_{i}\left( a_{1},a_{2},a_{3}\right) =0,
\end{equation}
and $K_{i},i=1,2,3$ are constants of integration satisfying the condition $%
\sum_{i=1}^{3}K_{i}=0$.

Therefore for an arbitrary Bianchi type geometry the anisotropy parameter
can be represented in the following exact form:
\begin{equation}
A=\frac{K^{2}+G^{2}+L}{V^{2}H^{2}}=9\frac{K^{2}+G^{2}+L}{\dot{V}^{2}},
\label{bian3}
\end{equation}
where
\begin{equation}
K^{2}=\frac{1}{3}\sum_{i=1}^{3}K_{i}^{2},
\end{equation}
\begin{equation}
G^{2}=\frac{1}{27}\sum_{i=1}^{3}\left[ \int \frac{\Delta F_{i}}{H}dV\right]
^{2},
\end{equation}
and
\begin{equation}
L=\frac{2}{9}\sum_{i=1}^{3}K_{i}\int \frac{\Delta F_{i}}{H}dV.
\end{equation}

Taking the time derivative of Eq. (\ref{bian3}), and changing the time
variable to $V$, it follows that in an arbitrary Bianchi type geometry the
anisotropy parameter satisfies a Bernoulli type first order differential
equation of the form
\begin{eqnarray}
\frac{dA}{dV} &=&\left[ \frac{d}{dV}\ln \left( K^{2}+G^{2}+L\right) \right]
A-  \nonumber \\
&&\frac{2V}{3\left( K^{2}+G^{2}+L\right) }\left[ F\left(
a_{1},a_{2},a_{3}\right) +\Lambda +%
\frac{k_{4}^{2}}{2}\left( \rho -p\right) -\frac{k_{5}^{4}}{12}\rho p+\frac{%
1}{3}k^{4}\frac{\mathcal{U}_{0}}{V^{4/3}}\right] A^{2},
\end{eqnarray}
with the general solution given by
\begin{equation}
A\left( V\right) =\frac{3\left( K^{2}+G^{2}+L\right) }{\int
V\left[ F\left( a_{1},a_{2},a_{3}\right) +\Lambda
+\frac{k_{4}^{2}}{2}\left( \rho -p\right)
-\,\frac{k_{5}^{4}}{12}\rho
p+\frac{1}{3}k^{4}\mathcal{U}_{0}V^{-4/3}\right] dV+C},
\end{equation}
where $C$ is an arbitrary constant of integration.

For a brane cosmological fluid obeying a linear barotropic
equation of state $p=\left(\gamma -1\right)\rho$, the anisotropy
parameter for arbitrary Bianchi type cosmological models is given
by
\begin{equation}
A\left( V\right) =\frac{3V^{2\gamma }\left( K^{2}+G^{2}+L\right)
}{V^{2\gamma }\int VF\left[ a_{1}\left( V\right) ,a_{2}\left(
V\right) ,a_{3}\left( V\right) \right] dV+\Lambda V^{2\left(
1+\gamma \right) }+k_{4}^{2}\rho
_{0}V^{2+\gamma }+\frac{k_{5}^{4}}{12}\rho _{0}^{2}V^{2}+k^{4}{\cal U}%
_{0}V^{2\left( 1/3+\gamma \right) }+CV^{2\gamma }},  \label{anf}
\end{equation}

We shall consider in the following the behavior of the anisotropy
parameter for arbitrary type Bianchi space-times, filled with a
linear barotropic cosmological fluid, obeying an equation of state
of the form $p=(\gamma -1)\rho $, $1\leq \gamma \leq 2$. For this
form of cosmological matter the conditions  $\rho ^{(eff)}>0$,
$\rho ^{(eff)}+p^{(eff)}>0$ and $\rho ^{(eff)}+3p^{(eff)}>0$ are
satisfied on the brane. Therefore the corresponding cosmological
models are singular \cite{ElHa73}. Since in most Bianchi types one
does not know any exact solution, in order to find the behavior of
the anisotropy at early times, it is necessary to use some
asymptotic solutions obtained, near the singularity, by
approximate methods, in the limit of small values of the time
parameter. In standard four-dimensional general relativity it is
concluded that for all Bianchi types there exists a Kasner-like
''vacuum phase'' near the singularity, that is, in general,
Einstein's vacuum equations are the first order approximation of
the equations with a nonvanishing matter term for $t\rightarrow
0$. This idea has been proposed a long time ago by Belinskii,
Lifshitz and Khalatnikov \cite{BeLiKh71}. The general
argument comes from the consideration of a Kasner type metric $%
ds^{2}=dt^{2}-t^{p_{1}}dx^{2}-t^{p_{2}}dy^{2}-t^{p_{3}}dz^{2}$, with $%
p_{i},i=1,2,3$ constants satisfying the conditions $\sum_{i=1}^{3}p_{i}=%
\sum_{i=1}^{3}p_{i}^{2}=1$. For a linear barotropic equation of state it
follows from the Bianchi identity (\ref{cons}) that energy density of the
matter behaves like $\rho \sim t^{-\gamma \left( p_{1}+p_{2}+p_{3}\right) }$%
. Therefore for asymptotes fulfilling the Kasner constraints the
matter term may be neglected in comparison with terms coming from
the geometric part of the field equations and containing the
second time derivatives. A Kasner type behavior near the
singularity for a Bianchi type IX brane world geometry has been
discussed in \cite{Co01a}

From Eq. (\ref{anf}) one can derive the behavior of the anisotropy parameter
near the singular state on the brane for all Bianchi types. We suppose that
the brane Universe starts its evolution from a singular state, with $%
a_{i}\left( 0\right) =0,i=1,2,3$. For $t>0$ and for an expanding geometry,
the scale factors are monotonically increasing functions of time, which for
small $t$ can be represented in the Kasner form
\begin{equation}
a_{i}\sim V^{m_{i}},i=1,2,3,
\end{equation}
with $m_{i}$ constants satisfying the conditions $0\leq m_{i}<1,i=1,2,3$ and
$\sum_{i=1}^{3}m_{i}=1$.

The existence of a such a representation for the scale factors near the
singular point on the brane has been explicitly proven, in the case of
Bianchi type I and V geometries, in \cite{ChHaMa01a}. The mean Hubble
parameter behaves near the singular state like $H\sim V^{-1}$.

With these assumptions it is easy to find the behavior of the integrals
involving the functions $F_{i}$ and $G_{i}$. For Bianchi class A models, $%
\int VF\left[ a_{1}\left( V\right) ,a_{2}\left( V\right) ,a_{3}\left(
V\right) \right] dV$ can be written, near the singular point, as a sum of
terms of the form $V^{4m_{i}},i=1,2,3$ and $V^{2m_{i}+2m_{j}},i\neq j$:
\begin{equation}
\int VF\left[ a_{1}\left( V\right) ,a_{2}\left( V\right) ,a_{3}\left(
V\right) \right] dV=\sum_{i=1}^{3}\alpha _{i}V^{4m_{i}}+\sum_{i\neq
j=1}^{3}\beta _{ij}V^{2m_{i}+2m_{j}},
\end{equation}
where $\alpha _{i},i=1,2,3$ and $\beta _{ij},i,j=1,2,3$ are constants.

Thus, for example, for the integral involving the function $F_{1}\left(
a_{1},a_{2},a_{3}\right) $ we obtain
\begin{equation}
\int VF_{1}\left( a_{1},a_{2},a_{3}\right) dV=\frac{c_{2}^{2}}{8m_{2}}%
V^{4m_{2}}-\frac{c_{2}c_{3}}{2\left( m_{2}+m_{3}\right) }V^{2\left(
m_{2}+m_{3}\right) }+\frac{c_{3}^{2}}{8m_{3}}V^{4m_{3}}-\frac{c_{1}^{2}}{%
8m_{1}}V^{4m_{1}}.
\end{equation}

For the other Bianchi types, the integral is a sum of terms of the form $%
V^{2-2m_{i}},i=1,2,3$ or $V^{2+2m_{3}-2m_{1}-2m_{2}}$ and cyclic permutation
of this term. Due to the relation $m_{1}+m_{2}+m_{3}=1$ the last expression
can be transformed to the form $V^{4m_{i}},i=1,2,3$. In the limit $%
V\rightarrow 0$, all these terms tend to zero.

The behavior of the function $G^{2}$ and $L$ can be analyzed in a similar
way, and one can easily show that $\lim_{V\rightarrow
0}G^{2}=\lim_{V\rightarrow 0}L=0$. Therefore for all Bianchi type
cosmological models the anisotropy parameter on the brane has the general
property $\lim_{V\rightarrow 0}A=0$. In the limit of large times, $%
V\rightarrow \infty $ and all Bianchi models (except type VII) isotropize,
with $A\rightarrow 0$.

\section{Discussions and final remarks}

The time evolution of the anisotropy parameter for a linear
barotropic fluid in the brane world model is very different from
the standard four-dimensional general relativistic case. In
homogeneous and anisotropic brane world models the Universe starts
from an isotropic state, with $A=0$. The anisotropy of the
Universe is increasing in time, and reaches a maximum value after
a finite time $t_{\max }$. For time intervals so that $t>t_{\max
}$, $A$ is a decreasing function of time which generally tends, in
the large time limit, to zero. In standard four-dimensional
general relativity the Universe starts its evolution from a
singular state with maximum anisotropy and reaches, for all
Bianchi models except type VII, an isotropic state for
$t\rightarrow \infty $ \cite{CoHa73}. For Bianchi type I and V
geometries the maximum value of the anisotropy parameter is
obtained for values of the volume scale factor $V$ satisfying the
equation
\begin{equation}
2\Lambda V+4\alpha ^{2}V^{1/3}+\left( 2-\gamma \right) k_{4}^{2}\rho
_{0}V^{1-\gamma }+\frac{k_{5}^{4}}{6}\rho _{0}^{2}\left( 1-\gamma \right)
V^{1-2\gamma }+\frac{2}{3}k^{4}{\cal U}_{0}V^{-1/3}=0,
\end{equation}
which follows from the condition $dA/dV=0$. In the case of the Bianchi type
I brane filled with a stiff cosmological fluid ($\gamma =2$), and for a
negligible small cosmological constant, $\Lambda =0$, the maximum value of
the volume scale factor is given by
\begin{equation}
V_{\max }=\left( \frac{k_{5}^{4}\rho _{0}^{2}}{4k^{4}{\cal U}_{0}}\right)
^{3/8}=\left( \frac{k_{4}^{4}\rho _{0}^{2}}{4{\cal U}_{0}}\right) ^{3/8}.
\end{equation}

The maximum value of the anisotropy parameter is given by
\begin{equation}
A_{\max }=A\left( V_{\max }\right) =\frac{3K^{2}}{k_{4}^{2}\rho _{0}+\sqrt{%
\frac{\rho _{0}}{8}}\frac{k_{5}^{4}{\cal U}_{0}^{3/4}}{k_{4}^{3}}+C}.
\end{equation}

The value of the cosmological time for which the $A$ reaches its maximum
value can be found from Eq. (\ref{dVH}), and it is given by
\begin{equation}
t_{\max }=\int_{0}^{V_{\max }}\frac{xdx}{\sqrt{Cx^{2}+3k^{4}{\cal U}%
_{0}x^{8/3}+k_{5}^{4}\rho _{0}^{2}/4}}.  \label{time}
\end{equation}

The integral in Eq. (\ref{time}) cannot be expressed in a simple analytical
form.

If in the early stages of evolution of a stiff ($\gamma =2$)
cosmological fluid on a Bianchi type I brane the dark radiation
term can be neglected, as being negligible small, ${\cal
U}_{0}\approx 0$, then the existence of a maximum of $A$ also
requires a non-zero cosmological constant, $\Lambda \neq 0$ For
${\cal U}_{0}=0$ the maximum value of the volume scale factor is
given by
\begin{equation}
V_{\max }=\left( \frac{k_{5}^{4}\rho _{0}^{2}}{12\Lambda }\right) ^{1/4},
\end{equation}
and the time necessary for the brane Universe to reach this state is
\begin{equation}
t_{\max }=\int_{0}^{V_{\max }}\frac{xdx}{\sqrt{3\Lambda x^{4}+Cx^{2}+\frac{%
k_{5}^{4}\rho _{0}^{2}}{4}}}=\frac{1}{2\sqrt{3\Lambda }}\ln \left( 1+\sqrt{%
\frac{2k_{5}^{2}\rho _{0}}{k_{5}^{2}\rho _{0}+\frac{C}{\sqrt{3\Lambda }}}}%
\right) .
\end{equation}

The maximum value of the anisotropy parameter is, in this case,
\begin{equation}
A_{\max }=A\left( V_{\max }\right) =\frac{3K^{2}}{\rho _{0}\left( k_{4}^{2}+%
\sqrt{\frac{\Lambda }{3}}k_{5}^{4}\right) +C}.
\end{equation}

This type of behavior of $A$, with an initial monotonically increasing
evolution from zero up to a maximum value, followed by a decrease to zero,
specific to brane world cosmological models, is a direct consequence of the
presence, in the gravitational field equations, of the term quadratic in the
energy density.

In the present paper we have considered in a systematic manner the
time evolution of the anisotropy parameter $A$ in the framework of
homogeneous, arbitrary Bianchi type brane world cosmological
models, with the matter content consisting of a perfect barotropic
cosmological fluid. For Bianchi type I and V exact representations
of $A$ can be obtained from the field equations and therefore the
initial behavior of $A$ can be explicitly derived from the study
of the exact representation near the singularity. In order to
obtain the form of $A$ at the initial moment of the cosmological
evolution for the other Bianchi types we have used the crucial
assumption that near the singularity the metric of the brane world
is of Kasner type. For a ''normal'' matter filled brane world,
satisfying a linear barotropic equation of state, the behavior of
$A$ is very different from the standard four-dimensional general
relativistic case. In this case the anisotropy parameter has the
remarkable property $\lim_{t\rightarrow 0}A(t)=\lim_{t\rightarrow
\infty }A(t)=0$, in sharp contrast to the SGR case. However, in
the case of pressureless dust, the initial values of the
anisotropy parameter are identical in both brane world and
standard four-dimensional general relativistic cosmological
models.

\section*{Acknowledgments}

The authors would like to thank to the two anonymous referees,
whose comments helped to significantly improve the manuscript.

\end{document}